\documentstyle[prd,aps,twocolumn,epsfig]{revtex}

\catcode`\@=11

\def\maketitle2{\par 
\begingroup
\let\cite\@bylinecite
\def\thefootnote{\fnsymbol{footnote}}%
\twocolumn[\@maketitle2\vskip2pc]%
\thispagestyle{plain}\@thanks
\endgroup
\def\thefootnote{\arabic{footnote}}%
\setcounter{footnote}{0}%
\let\maketitle2\relax \let\@maketitle2\relax
\let\@thanks\relax \let\@authoraddress\relax \let\@title\relax
\let\@date\relax \let\thanks\relax \let\@abstract\relax 
\let\@pacs\relax}

\def\abstract#1{\gdef\@abstract{{\par 
\bgroup
\ifdim\prevdepth=-1000pt \prevdepth0pt\fi
\hsize\columnwidth
\dimen0=-\prevdepth \advance\dimen0 by17.5pt \nointerlineskip
\small\vrule width 0pt height\dimen0 \relax}{~~}#1\egroup}}

\def\pacs#1{\gdef\@pacs{{\par 
\bgroup
\hsize\columnwidth \parindent0pt
\ifdim\prevdepth=-1000pt \prevdepth0pt\fi
\dimen0=-\prevdepth \advance\dimen0 by20pt\nointerlineskip
\egroup} PACS numbers:~#1}}

\def\@maketitle2{
\@preprint
\@title
\ifdim\prevdepth=-1000pt \prevdepth0pt\fi
\@authoraddress
\@date
\begin{list}{}{\leftmargin=0.10753\textwidth \rightmargin=\leftmargin
\itemsep=1pc\partopsep=-1pc}
\item\@abstract
\item\@pacs
\end{list}
}

\catcode`\@=12

\begin{document}

\title{Superinflation, quintessence, and nonsingular cosmologies}
\author {E. Gunzig$^{1,2}$, A. Saa$^{1,3,4}$, L. Brenig$^{1,5}$,
V. Faraoni$^{1,6}$, T.M. Rocha Filho$^{7}$ and A.
Figueiredo$^{7}$.}
\address {$^1$ RggR, Universit\'e Libre de Bruxelles, 
CP 231, 1050 Bruxelles, Belgium} 
\address{$^2$ Instituts Internationaux de Chimie et de Physique Solvay,
CP231, 1050 Bruxelles, Belgium}
\address{$^3$ Departament de F\'\i sica Fonamental, Universitat de Barcelona,
Av. Diagonal 647, 08028 Barcelona, Spain }
\address{$^4$ Departamento de Matem\'atica Aplicada, IMECC--UNICAMP, CP6065,
13081-970 Campinas, SP, Brazil }
\address{$^5$ Service de Physique Statistique, Universit\'e Libre de
Bruxelles, CP231, 1050 Bruxelles, Belgium}
\address{$^6$ INFN-Laboratori Nazionali di Frascati, Box 13, 00044
Frascati, Roma, Italy}
\address{$^7$ Instituto de F\'\i sica, Universidade de Brasilia, 70.910-900
Brasilia, DF, Brazil}

\date{\today}

\abstract
{The dynamics of a universe dominated by a self-interacting nonminimally
coupled scalar field are considered. The structure of the phase
space and complete phase portraits are given. New dynamical behaviors
include superinflation ($\dot{H}>0$), avoidance of big bang singularities
through classical birth of the universe, and
spontaneous entry
into and exit from inflation. This model is  promising for describing 
quintessence as a nonminimally coupled scalar field.}


\maketitle2
\narrowtext
A dynamical system approach to a self-consistent nonsingular cosmological
history is presented in the framework of the classical Einstein equations 
with a nonminimally coupled scalar field.   
The description of the matter content of the cosmos with a single scalar 
is appropriate during
important epochs of the history of the universe \cite{KolbTurner}. A
crucial ingredient of
the physics of scalar
fields in curved spaces is their nonminimal coupling to the Ricci
scalar $R$ of spacetime, which is required by first loop corrections
\cite{loop},  
by specific particle theories \cite{preprint}, 
and by scale-invariance arguments at the classical level
\cite{CCJ}. It is well known that
nonminimal coupling dictates the success or failure of inflationary
models \cite{Abbott}; more generally, it turns out to
strongly affect the cosmic dynamics, which is qualitatively richer than 
in the minimally coupled case.  We show, indeed, that nonminimal coupling
leads to new dynamical behaviors, such as a regime that we propose to call
{\em superinflation} ($\dot{H}>0$, which cannot be achieved with minimal
coupling \cite{LiddleParsonsBarrow}), spontaneous entry into
and exit from inflation, with or without a cosmological constant,
and a possible model for quintessence. 
Spontaneous superinflation provides a classical
alternative, which is
impossible with minimal coupling, to semiclassical\cite{GunzigNardone,IJTP1}
and quantum\cite{Vilenkin} birth of the universe.

We consider the nonminimally coupled theory described by the action
\begin{equation}   \label{action}
S=\frac{1}{2} \int d^4x\sqrt{-g}\left(-\, \frac{R}{\kappa } 
+g^{\mu \nu}\partial_\mu \psi \partial_\nu \psi 
-2V +\xi R\psi ^2  \right) \; ,
\end{equation}
where $\kappa \equiv 8\pi G$ ($G$ being Newton's constant), $\xi $ is the
nonminimal coupling constant, and a cosmological constant $\Lambda$, if
present, is incorporated in the scalar field potential $V(\psi )$.
We use the full conserved scalar field stress-energy tensor 
\begin{eqnarray}   \label{stressenergy}
T_{\mu\nu}& =&  \partial_{\mu}\psi \partial_{\nu} \psi-\xi \left(
\nabla_{\mu}\nabla_{\nu}
-g_{\mu\nu} \Box \right)( \psi^2 )+\xi G_{\mu\nu} \psi^2 \nonumber\\
& & -\frac{1}{2} g_{\mu\nu} \left( \partial_{\alpha} \psi 
\partial^{\alpha} \psi -2V( \psi) \right) 
\end{eqnarray}
(where $G_{\mu\nu}$ is the Einstein tensor), thereby 
avoiding the widespread effective coupling
$\kappa_{\rm eff}=\kappa \left( 1-\kappa \xi\psi^2 \right)^{-1}$
in the Einstein equations
$G_{\mu\nu} =\kappa T_{\mu\nu}$.  We
study the dynamics of a spatially flat Friedmann-Robertson-Walker universe
with line element $ds^2=d\tau^2-a^2( \tau) \left(  dx^2+dy^2+dz^2
\right)$. This yields the trace equation $R=-\kappa \left(
\sigma-3p\right)$, the energy constraint $3H^2=\kappa\sigma$ (which
guarantees that the energy density $\sigma \geq 0$), and the
Klein-Gordon equation, respectively
\begin{eqnarray}  \label{trace}
& & 6\left[ 1 -\xi \left( 1- 6\xi \right) \kappa \psi^2
\right] \left( \dot{H} +2H^2 \right) 
-\kappa \left( 6\xi -1 \right) \dot{\psi}^2  \nonumber \\ 
& & - 4 \kappa V  + 6\kappa \xi \psi \frac{dV}{d\psi} = 0 \; , 
\end{eqnarray}
\begin{equation} \label{Hamiltonianconstraint}
\frac{\kappa}{2}\,\dot{\psi}^2 + 6\xi\kappa H\psi\dot{\psi}
- 3H^2 \left( 1-\kappa \xi \psi^2 \right) + \kappa  V =0 \, ,
\end{equation}
\begin{equation}   \label{KleinGordon}
\ddot{\psi}+3H\dot{\psi}+\xi R \psi +\frac{dV}{d\psi} =0 \; .
\end{equation}
The (time-dependent) equation of state of the $\psi$ field, rather than
being imposed {\em
a priori}, follows self-consistently from the dynamics.
The system~(\ref{trace})-(\ref{KleinGordon}) reduces to a 
two-dimensional set of first order equations for 
$\psi$ and $H\equiv \dot{a}/a$  (note that this dimensional reduction is
not possible for spatially curved universes \cite{IJTP1,ALO}). 
One has
\begin{equation}  \label{psidot}
\dot{\psi}=-6 \xi H \psi \pm\frac{1}{2\kappa} \sqrt{{\cal G} ( H, \psi
)}
\; ,
\end{equation}
\begin{eqnarray} \label{Hdot}
\dot{H}&=& \frac{1}{1+ \kappa \xi ( 6\xi -1 ) \psi^2 } 
\left[ 3 \left( 2\xi -1 \right) H^2 
 \right.\nonumber\\ 
&&   + 3 \xi ( 6\xi -1 )
( 4\xi -1)\kappa H^2 \psi^2 \mp \xi
( 6\xi -1) H \psi \sqrt{{\cal G}} \nonumber\\
&& \left. +\left( 1-2\xi \right) \kappa V ( \psi ) -\kappa \xi \psi
\frac{dV}{d\psi} \right] \; ,    
\end{eqnarray}
where
\begin{equation} \label{G}
{\cal G} ( H, \psi ) = 8 \kappa^2 \left[ \frac{3H^2}{ \kappa} -V ( \psi )
+3 
\xi ( 6\xi -1 ) H^2 \psi^2  \right] \; .
\end{equation}
Due to the energy constraint (\ref{Hamiltonianconstraint}), the
trajectories are restricted 
to a two-dimensional manifold $\Sigma$ in the three-dimensional
$\left( H, \psi, \dot{\psi} \right) $ phase space, possibly with ``holes''
(dynamically forbidden regions) 
corresponding to ${\cal G}(H,\psi)<0 $ (cf. Eq.~(\ref{psidot})). $\Sigma$
is composed of two sheets corresponding to the positive or negative sign
in Eq.~(\ref{psidot}). The
two sheets  smoothly
join on the boundary ${\cal G}=0$ of the dynamically
forbidden region. 

In the following, for simplicity, we project the dynamics of the phase
space onto the $(H, \psi )$ plane, but the true  
nature of $\Sigma$ should always be kept in mind. 
We now restrict to the widely used potential
\begin{equation} \label{potential}
V( \psi ) = \frac{3\alpha}{\kappa} \psi^2 -\frac{\Omega}{4} \psi^4 -
\frac{9\omega}{\kappa^2} \;,
\end{equation}
consisting of a mass term, a quartic self-coupling and, possibly, a
cosmological constant term. For consistency with
previous works \cite{IJTP1} we use the dimensionless
symbols $\alpha \equiv \kappa
m^2/6$ ($m$ being the scalar field mass) and $\omega \equiv
-\kappa^2\Lambda/9 $.

The fixed points of the system (\ref{trace})-(\ref{KleinGordon}) are
the de Sitter solutions with constant scalar field
\begin{equation}   \label{fixedpoints}  
H^2_0=\frac{ 3(\alpha^2-\Omega\omega)}{\kappa ( \Omega - 6\xi\alpha )} \;,
\;\;
\psi^2_0=\frac{ 6(\alpha-6\xi\omega)}{\kappa ( \Omega - 6\xi\alpha )} 
\end{equation}
$\left( \Omega\neq 6\alpha\xi \right)$, and 
$\left( H,\psi \right)=\left( \pm \sqrt{-3\omega/\kappa}, 0 \right)$.
The fixed points  (\ref{fixedpoints})  exist also
for $\omega=0$, due to the presence of the matter field $\psi$ (in this
case, the two points $\left( \pm \sqrt{-3\omega/\kappa}, 0 \right)$
collapse into the Minkowski space fixed point $(0,0$). We only
present here the case of conformal coupling, $\xi=1/6$.

The function 
\begin{equation} \label{lyapunov}
L( \psi, \dot{\psi})= \frac{1}{2} \dot{\psi}^2 + \frac{\alpha}{4} \,
\psi^4 - \frac{3\omega}{\kappa} \, \psi^2 + V( \psi) 
\end{equation}
is such that $dL/dt =-3H\dot{\psi}^2$ along the trajectories. For $H>0$,
$L$ is a Lyapunov function  in a region containing 
the origin; the solutions are then confined  by the closed lines of constant
$L$, 
implying asymptotic convergence to the fixed points on the $H$ axis. 
This behavior is confirmed by exhaustive numerical simulations
reported in the following.
We first exclude a cosmological constant
by setting $\omega=0$. The phase portrait qualitatively differs according
to
the ratio $\Omega/\alpha$.

\begin{figure}[ht]
\epsfxsize=8.0cm
\epsfbox{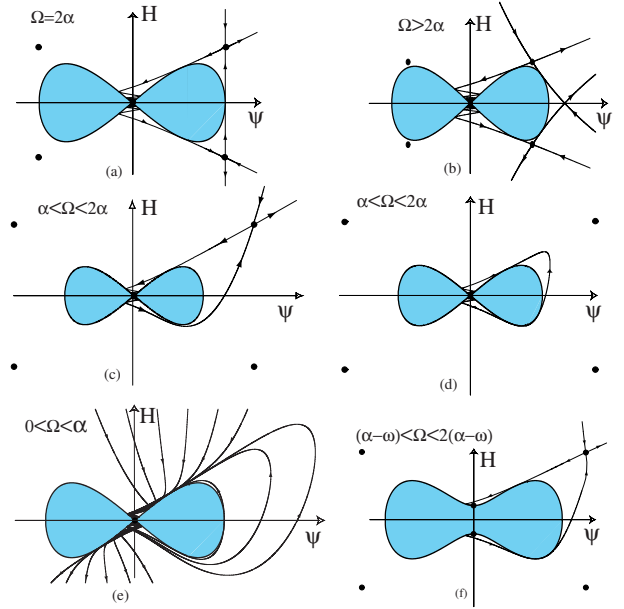}
\vspace{0.5cm}
\caption{Qualitative phase portrait for the system  
(\ref{trace})-(\ref{KleinGordon}). Shadowed regions correspond to the
dynamically forbidden regions (${\cal G}(H,\psi)<0 $, cf. Eq.~(\ref{psidot})),
and the dots to the fixed points.
Figures a-e, were obtained by using $\Omega/\alpha$, respectively,
2, 5, 3/2, 3/2, and 1/2, and $\omega=0$. Figure f corresponds to
the case $\omega=-1/10$ and $\Omega/(\alpha-\omega)=3/2$.}
\end{figure}

{\bf The case $\Omega=2\alpha$:} The Minkowski space   $\left( H, \psi,
\dot{\psi} \right)=\left( 0, 0, 0
\right) $ is a fixed point, attractive for $H>0$ and repulsive for $H<0$;
the projections of the de Sitter spaces 
$\left( \pm H_0, \pm \psi_0, 0 \right) $ are saddle points, i.e. they
possess attractive and repulsive eigendirections in the phase space
as shown by the arrows in (Fig.~1a).
The new kind of solutions, only present in this case
$\Omega=2\alpha$ and unavoidably 
missed in the approach using $\kappa_{\rm eff}(\tau)$,
\begin{equation} \label{heteroclinic}
H( \tau ) =\sqrt{\frac{C}{2}} \tanh \left( \sqrt{2C}\, \tau \right)
\;,\;\;\;\;\;\;
\psi=\pm \psi_0 \equiv \pm \sqrt{\frac{6}{\kappa}} 
\end{equation}
(where $C=\dot{H}+2H^2 = -R/6 $ is constant), corresponds to heteroclinic
straight
lines connecting de Sitter fixed points, starting along
the repulsive eigendirection of one of them and ending along the
attractive eigendirection of the other (Fig.~1a). They are
tangent to the boundary of the forbidden regions at  
$\left( H, \psi \right) =\left( 0, \pm\psi_0 \right) $.
For $\left| H \right| >\sqrt{C/2}$, another straight line solution is
obtained from the general form
\begin{equation} \label{exact}
H( \tau)=\sqrt{\frac{C}{2}}\, \frac{w_1 {\mbox e}^{\sqrt{C/2}\tau}-
w_2 {\mbox e}^{-\sqrt{C/2}\tau}}
{w_1 {\mbox e}^{\sqrt{C/2}\tau}+
w_2 {\mbox e}^{-\sqrt{C/2}\tau}}\; , \psi=\psi_0 \; ,
\end{equation}
where $w_1$ and $w_2$ are integration constants.
The nonsingular solutions (\ref{heteroclinic}) connect a
contracting ($\tau \rightarrow -\infty$) de Sitter regime to an
expanding one ($\tau \rightarrow +\infty$) and passes through
a minimum
nonvanishing value of the scale factor ($\tau=0$).

In addition to these straight lines, there exist 
  other heteroclinic solutions: one starting at $\left(
H,\psi \right)=( 0,0) $ and ending at  
$( -H_0, \psi_0) $, and another one from 
$\left( H_0,\psi_0 \right)$ to $( 0,0) $. A third solution starts at
the expanding de Sitter point and goes to infinity, while another one
comes from infinity and arrives to the contracting de Sitter point. The
phase portrait is symmetric about the origin.

Near the Minkowski fixed point $\left(
0,0,0 \right)$, numerical analysis confirms the peculiar behavior
suggested
by the Lyapunov function: orbits approaching this point with positive
$H$ are attracted to it, bouncing back and forth infinitely many times off
the
${\cal G}=0$ boundary in the $\left( H, \psi \right) $ projection
(Fig.~1a). In the space $\left( H,
\psi,
\dot{\psi} \right) $ these orbits are seen to spiral down on a cone
towards its apex at the origin.  
The cone results from the union of the two sheets in the vicinity of the
origin. Along the spiral, the orbit passes almost periodically from one
sheet to the other with period $ \tau_{\rm bounce}=2\pi/m$ 
($\tau_{\rm bounce}$ is
obtained by an asymptotic analysis of 
the system (\ref{trace})-(\ref{KleinGordon}). Typically, 
after a few bounces, the period coincides with
$\tau_{\rm bounce} $ with good accuracy). A similar 
behavior for $\Omega <0$ was reported in the earlier numerical analysis
of \cite{ALO}, but  using the effective coupling $\kappa_{\rm eff}( \tau)$ and
the variables $\psi$ and $\dot{\psi}$.

In the $H<0$ half-plane, the situation is reversed: orbits starting with
$H<0$ are repelled by the origin and depart from it bouncing
off the ${\cal G}=0$ boundary.

{\bf The case $\Omega> 2\alpha$:} the situation (Fig.~1b) is analogous to
the previous one, but now the straight heteroclinic solutions are
missing, and are replaced by the solution starting at the contracting
de Sitter fixed point and escaping to infinity, and by the solution coming
from infinity and arriving to the expanding de Sitter point. The
two-sheeted structure of $\Sigma$ implies that no actual intersections
occur between different orbits in Fig.~1b, which live in different
sheets but are projected in the same plane.

{\bf The case $\alpha < \Omega< 2 \alpha $:} as shown in Fig.~1c, there
are no straight heteroclinic lines but
new interesting features emerge. A new heteroclinic solution appears
starting from the origin and ending in the expanding de Sitter fixed
point. As in the previous case, the quadrant $\psi>0$, $H<0$ is obtained
from the $\psi>0$, $H>0$ one by reflection about the $\psi$-axis and
time-reversal. 

The crucial feature of this case is the appearance of a dense set of
homoclinic solutions (Fig.~1d) departing from the origin with negative $H$
and
returning to it with positive $H$, going around the forbidden region.
Superinflation plays a central role along these orbits; only a regime with
$\dot{H}>0$ permits the smooth transition from an initial contracting
($H<0$) phase to an expanding one ($H>0$). This transition occurs at the
nonvanishing minimum of the scale factor. The behavior of this family of
homoclinics, as well as of the other solutions, is universal: they rapidly
converge in the spiraling region near the origin, irrespective of 
initial conditions. As all of these homoclinics originate around Minkowski
fixed point
 due to its instability with respect to perturbations with $H<0$,
this 
dynamics constitute a classical alternative to the previously proposed
semiclassical birth of the universe from empty space
\cite{GunzigNardone,IJTP1}.

{\bf The case $0<\Omega< \alpha$:} the fixed points (\ref{fixedpoints})
disappear and the only bounded solutions are the
homoclinics
associated with the origin (see Fig.~1e). This situation is therefore the
most favorable for the classical spontaneous exit from empty Minkowski
fixed point.

Analogous results hold also with a small and positive 
cosmological constant $\Lambda$ (see, for instance,
Fig.~1f), with the phase portrait being classified according to
$\Omega / ( \alpha-\omega )$, but the fixed point $(0,0,0)$ of the
$\omega=0$ case splits into two de Sitter fixed points with memory of
the previous stability properties. Now, the approximate period between two
consecutive bounces is 
\begin{equation}
\tau_{\rm bounce}=\frac{2\pi}{\sqrt{m^2-\frac{1}{12}\kappa\Lambda } } \; .
\end{equation}

\begin{figure}[ht]
\epsfxsize=8.0cm
\epsfbox{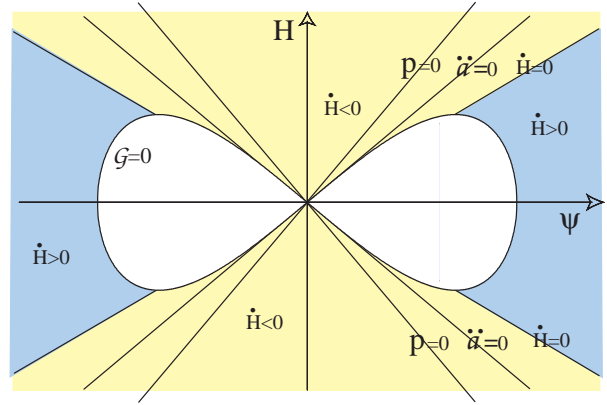}
\vspace{0.5cm}
\caption{The plane $(H,\psi)$ for the
$\omega=0$ case and the equation of state. The darkest
region corresponds to the superinflation regime ($\dot{H}>0$).
In the region $\dot{H}<0$
and $H>0$, the
equation of state associated with the bouncing solution $\psi$
passes periodically through  
radiation domination (crossing the $H$-axis),
matter domination ($p=0$), and re-acceleration (between ${\cal G}=0$
and $\ddot{a}=0$ lines). Asymptotically, as $\tau
\rightarrow +\infty$, 
the universe becomes matter dominated $a( \tau) \propto \tau^{2/3}$ and
tends to infinite dilution.}
\end{figure}

Returning to the $\omega=0$ case,
the $ \left( H, \psi \right)$ plane is divided into sectors by the
straight lines $H=\pm \sqrt{\alpha/2} \, \psi$, 
$H=\pm \sqrt{\alpha} \, \psi$
$H=\pm \sqrt{2\alpha} \, \psi$ corresponding, respectively, to
$\dot{H}=0$, $\ddot{a}=0$, and pressure $p=0$ (the $H$-axis  corresponds
to $p=\sigma/3$). The lines  $H=\pm \sqrt{\alpha/2} \, \psi$ mark the
transition between inflationary ($\ddot{a}>0$ and $\dot{H} \leq 0$) and
superinflationary ($\dot{H}>0$)  regimes. The lines 
$H=\pm \sqrt{\alpha} \, \psi$ divide regions corresponding to inflation
and to decelerated expansion, while 
$H=\pm \sqrt{2\alpha} \, \psi$ divide regions of positive and negative
pressures (Fig.~2). The crucial condition for superinflation to occur is
that  the line $\dot{H}=0$ (or parts of it) belong
to the dynamically accessible region ${\cal G}\geq 0$ of the $\left( H,
\psi \right) $ plane. This implies that, for an arbitrary potential $V(
\psi)$,
superinflation corresponds to $\psi dV/d\psi \leq 0 $. 
For our particular potential
(\ref{potential}), this requires $\Omega>0$. 
The superinflationary behavior occurs only once along each homoclinic and
brings the solution from the primordial Minkowski fixed point 
neighborhood to the
succession of eras corresponding to different equations of state (during
each bounce), towards infinite dilution and equation of state $p=0$. Indeed,
 asymptotic analysis for $\tau \rightarrow +\infty$ and for any
value of $\alpha$ and $\Omega$ (with $\omega=0$) shows that the scale
factor $a( \tau ) $ exhibits oscillations of concavity corresponding to
accelerated and decelerated epochs. These oscillations are damped as $\tau
\rightarrow +\infty$; in this regime, $a( \tau) \propto \tau^{2/3}$ and
the universe becomes matter dominated.

While it is not claimed here that the evolution of our universe is
modeled by an entire orbit of the system 
(\ref{trace})-(\ref{KleinGordon}) on the accessible manifold  
$\Sigma$, the application to specific eras of the cosmological history is
intriguing.
Indeed, in the bounces reported above (during which $\dot{H}<0$), one
encounters, respectively, radiation domination crossing the $H$-axis,
matter domination ($p=0$), acceleration (a possible quintessence model?)
until the next bounce in the $\left( H, \psi \right)$ projection, where
this sequence is reversed.
If we identify one period as our cosmological history, then the reported
accelerated expansion of the universe today \cite{SN} suggests to locate
our epoch in the sector between the line $H=\sqrt{\alpha} \, \psi$ and the
${\cal G}=0$ boundary. The identification of the age of the universe
($\sim 10^{17}$~s) with
$\tau_{\rm bounce}$ would then yield the scalar field mass $m \simeq
10^{-13}$~eV, which is suggestive of an axion \cite{KolbTurner} or of an 
ultralight pseudo-Goldstone boson (that the quintessence field should be
very light was already suggested \cite{Binetruy}).

Although our simplified model is not a quintessential scenario, its
features are promising for the description of quintessence as a
nonminimally coupled scalar field. 
Furthermore, the main properties of the model presented here
remain unaltered for a large class of potentials $V(\psi)$.
Finally, the fact that the dynamics of
the system
(\ref{trace})-(\ref{KleinGordon}) are confined to the two-dimensional
smooth manifold $\Sigma$ indicates 
the absence of chaos, a conclusion supported by
further analysis. This is not the case, a priori, of a spatially
curved universe and/or of multiple matter fields.

We acknowledge the Centre de Calcul Symbolique sur Ordinateur for
the use of computer facilities, and financial support from the EEC
(grant HPHA-CT-2000-00015), from OLAM, Fondation pour la Recherche
Fondamentale (Brussels), from FAPESP (S\~ao Paulo, Brazil), and from the
ESPRIT Working Group CATHODE.

\end{document}